\documentstyle[12pt]{article}
\input epsf
\newcommand{\nc}{\newcommand}
\nc{\bg}{B. Grz\c{a}dkowski}
\nc{\non}{\nonumber}
\def\gesim{\lower0.5ex\hbox{$\:\buildrel >\over\sim\:$}} 
\def\lesim{\lower0.5ex\hbox{$\:\buildrel <\over\sim\:$}} 
\nc{\prd}[3]{{\it Phys.\ Rev.}\ {{\bf D{#1}} (#2), #3}}
\nc{\prl}[3]{{\it Phys.\ Rev.\ Lett.}\ {{\bf {#1}} (#2), #3}}
\nc{\plb}[3]{{\it Phys.\ Lett.}\ {{\bf B{#1}} (#2), #3}}
\nc{\npb}[3]{{\it Nucl.\ Phys.}\ {{\bf B{#1}} (#2), #3}}
\nc{\ptp}[3]{{\it Prog.\ Theor.\ Phys.}\ {{\bf {#1}} (#2), #3}}
\nc{\zfp}[3]{{\it Z.\ Phys.}\ {{\bf C{#1}} (#2), #3}}
\nc{\mpla}[3]{{\it Mod.\ Phys.\ Lett.}\ {{\bf A{#1}} (#2), #3}}
\nc{\rmp}[3]{{\it Rev.\ Mod.\ Phys.}\ {{\bf {#1}} (#2), #3}}
\nc{\ijmpa}[3]{{\it Int.\ J.\ of\ Mod.\ Phys.}\
               {{\bf A{#1}} (#2), #3}}
\nc{\eps}{\epsilon}
\nc{\psp}{Ps_+}        \nc{\psm}{Ps_-}        \nc{\lsp}{ls_+}
\nc{\lsm}{ls_-}        \nc{\sss}{s_+s_-}      \nc{\m}{m_t}
\nc{\mq}{m_t^2}        \nc{\mr}{\frac{1}{\m}} \nc{\av}{A_{\gamma}}
\nc{\bv}{B_{\gamma}}   \nc{\az}{A_Z}          \nc{\bz}{B_Z}
\nc{\avs}{A_{\gamma}^2}\nc{\azs}{A_Z^2}       \nc{\bzs}{B_Z^2}
\nc{\dav}{\delta \! A_{\gamma}}   \nc{\dbv}{\delta \! B_{\gamma}}
\nc{\dcv}{\delta C_{\gamma}}      \nc{\ddv}{\delta \! D_{\gamma}}
\nc{\daz}{\delta \! A_Z}          \nc{\dbz}{\delta \! B_Z}
\nc{\dcz}{\delta C_Z}             \nc{\ddz}{\delta \! D_Z}
\nc{\dev}{\delta \! E_{\gamma}}   \nc{\dez}{\delta \! E_Z}
\nc{\dfv}{\delta \! F_{\gamma}}   \nc{\dfz}{\delta \! F_Z}
\nc{\rdav}{{\rm Re}(\delta \! A_{\gamma}) \:}
\nc{\rdbv}{{\rm Re}(\delta \! B_{\gamma}) \:}
\nc{\rdcv}{{\rm Re}(\delta C_{\gamma}) \:}
\nc{\rddv}{{\rm Re}(\delta \! D_{\gamma}) \:}
\nc{\rdaz}{{\rm Re}(\delta \! A_Z) \:}
\nc{\rdbz}{{\rm Re}(\delta \! B_Z) \:}
\nc{\rdcz}{{\rm Re}(\delta C_Z) \:}
\nc{\rddz}{{\rm Re}(\delta \! D_Z) \:}
\nc{\idav}{{\rm Im}(\delta \! A_{\gamma}) \:}
\nc{\idbv}{{\rm Im}(\delta \! B_{\gamma}) \:}
\nc{\idcv}{{\rm Im}(\delta C_{\gamma}) \:}
\nc{\iddv}{{\rm Im}(\delta \! D_{\gamma}) \:}
\nc{\idaz}{{\rm Im}(\delta \! A_Z) \:}
\nc{\idbz}{{\rm Im}(\delta \! B_Z) \:}
\nc{\idcz}{{\rm Im}(\delta C_Z) \:}
\nc{\iddz}{{\rm Im}(\delta \! D_Z) \:}
\nc{\cz}{(1+v_e^2)d\:\!'^2}         \nc{\ci}{v_ed\:\!'}
\nc{\ccz}{v_ed\:\!'^2}              \nc{\cci}{d\:\!'}
\nc{\lspace}{\;\;\;\;\;\;\;\;\;\;}  \nc{\llspace}{\lspace \lspace}

\nc{\beq}{\begin{equation}}   \nc{\eeq}{\end{equation}}
\nc{\bea}{\begin{eqnarray}}   \nc{\eea}{\end{eqnarray}}
\nc{\baa}{\begin{array}}      \nc{\eaa}{\end{array}}
\nc{\bit}{\begin{itemize}}    \nc{\eit}{\end{itemize}}
\nc{\ben}{\begin{enumerate}}  \nc{\een}{\end{enumerate}}
\nc{\bce}{\begin{center}}     \nc{\ece}{\end{center}}
\begin{document}
\pagestyle{empty} \setlength{\footskip}{2.0cm}
\setlength{\oddsidemargin}{0.5cm} \setlength{\evensidemargin}{0.5cm}
\renewcommand{\thepage}{-- \arabic{page} --}
\def\mib#1{\mbox{\boldmath $#1$}}
\def\bra#1{\langle #1 |}      \def\ket#1{|#1\rangle}
\def\vev#1{\langle #1\rangle} \def\dps{\displaystyle}
\def\sla#1{\mbox{$#1\hspace*{-0.17cm}\scriptstyle{/}\:$}}
\renewcommand{\thepage}{-- \arabic{page} --}
   \def\thebibliography#1{\centerline{REFERENCES}
     \list{[\arabic{enumi}]}{\settowidth\labelwidth{[#1]}\leftmargin
     \labelwidth\advance\leftmargin\labelsep\usecounter{enumi}}
     \def\newblock{\hskip .11em plus .33em minus -.07em}\sloppy
     \clubpenalty4000\widowpenalty4000\sfcode`\.=1000\relax}\let
     \endthebibliography=\endlist
   \def\sec#1{\addtocounter{section}{1}\section*{\hspace*{-0.72cm}
     \normalsize\bf\arabic{section}.$\;$#1}\vspace*{-0.3cm}}
\vspace*{-1.6cm}\noindent
\hspace*{11.cm}TOKUSHIMA 98-04\\
\hspace*{11.cm}(hep-ph/9810490)\\

\vspace*{1cm}

\begin{center}
{\large\bf Remark on $\mib{C}\!\mib{P}$-violating Polarization
Asymmetry}

\vskip 0.15cm
{\large\bf of $\mib{t}\bar{\mib{t}}$ via Anomalous Couplings to
$\mib{\gamma}$ and $\mib{Z}$}
\end{center}

\vspace*{1.5cm}
\begin{center}
\renewcommand{\thefootnote}{\alph{footnote})}
{\sc Zenr\=o HIOKI}$\,^{1),}$\footnote{E-mail address:
\tt hioki@ias.tokushima-u.ac.jp} and 
{\sc Kazumasa OHKUMA}$\,^{2),}$\footnote{E-mail address:
\tt ohkuma@radix.h.kobe-u.ac.jp}

\vspace{0.5cm}
$1)$ {\sl Institute of Theoretical Physics,
\ University of Tokushima}\\ {\sl Tokushima 770-8502, JAPAN}

\vspace{0.3cm}
$2)$ {\sl Graduate School of Science and Technology,
Kobe University}\\ {\sl Nada, Kobe 657-8501, JAPAN}
\end{center}

\vspace*{2.5cm}
\centerline{ABSTRACT}

\vspace*{0.4cm}
\baselineskip=20pt plus 0.1pt minus 0.1pt
We calculate the $C\!P$-violating polarization asymmetry of $t
\bar{t}$, $\delta\equiv[\:\sigma(e\bar{e}\to t(-)\bar{t}(-))-
\sigma(e\bar{e}\to t(+)\bar{t}(+))\:]/\sigma(e\bar{e}\to t\bar{t})$,
for the most general $t\bar{t}\gamma/Z$ couplings without dropping
any non-standard contribution. We find that one term which is usually
neglected increases with $s$ and will eventually become
non-negligible at very high energy.
\vfill
\newpage
\renewcommand{\thefootnote}{\sharp\arabic{footnote}}
\pagestyle{plain} \setcounter{footnote}{0}
\baselineskip=21.0pt plus 0.2pt minus 0.1pt

The Standard Model (SM) of electroweak interaction has been very
successful in describing various experimental data up to the scale of
$O(M_{W,Z})$. Still many people believe in the existence of new
physics at higher-energy scale which reduces the number of the free
parameters in the SM. Since the top-quark couplings have not been
studied in detail yet, there could be a room for new physics in them.
In the near future, experiments of top-quark-pair production via $e^+
e^-$ annihilation are expected to become possible at Next Linear
Collider (NLC). Indeed, NLC will be a powerful tool for new-physics
search, and provide us useful data for probing the top-quark
couplings to the photon and $Z$ boson.

One of such studies will be a measurement of $C\!P$ violation through
possible anomalous couplings. Since $t\bar{t}$ pairs are produced
mainly through the vector-boson ($\gamma/Z$) exchange, the helicities
of $t\bar{t}$ would be only ($+-$) or ($-+$) if top-quark mass were
much smaller than $\sqrt{s}$. However, because the observed top mass
is $173.9\pm 5.2$ GeV \cite{mt}, we can also expect ($++$) and ($--$)
combinations. This enables us to study $C\!P$ violation, because 
$$
\hat{C} \hat{P}| \mp \mp \rangle =| \pm \pm \rangle
$$
and therefore $C\!P$ violation can be measured through the asymmetry
\begin{equation}
\delta \equiv \frac{N(--) - N(++)}{N(all)} \label{eq:delta}
\end{equation}
where $N(all) \equiv N(++)+N(-+)+N(+-)+N(--)$.

On this theme have appeared so far many papers
\cite{SP}--\cite{Papers}. In those papers, products of non-standard
parameters were usually neglected and only interference between the
SM and non-SM terms was taken into account under the assumption that
non-SM effects are tiny. At much higher energy, however, such
neglected terms might come to play an important role. Therefore, in
this short note we calculate the cross section of polarized top-pair
productions via $e^+e^-$ starting from the most general form of top
interaction and keeping those which are usually neglected. Then we
study their effects on the above $C\!P$-violating asymmetry
numerically.

In our calculation, we will assume the following $t\bar{t}\gamma/Z$
couplings:
\begin{equation}
{\mit\Gamma}_{vt\bar{t}}^{\mu}=
\frac{g}{2}\,\bar{u}(p_t)\,\Bigl[\,\gamma^\mu \{A_v+\delta\!A_v
-(B_v+\delta\!B_v) \gamma_5 \}
+\frac{(p_t-p_{\bar{t}})^\mu}{2m_t}(\delta C_v-\delta\!D_v\gamma_5)
\,\Bigr]\,v(p_{\bar{t}}),\ \label{ff}
\end{equation}
where $g$ denotes the $SU(2)$ gauge coupling constant, $v=\gamma,Z$,
and
\[
\av=\frac43 \sin\theta_W,\ \ \bv=0,\ \ \az=\frac{v_t}{2\cos\theta_W},
\ \ \bz=\frac{1}{2\cos\theta_W}
\]
with $v_t\equiv 1-(8/3)\sin^2\theta_W$. $\delta\!A_v$, $\delta\!B_v$,
$\delta C_v$ and $\delta\!D_v$ are parameters expressing non-standard
effects.\footnote{In fact, the most general form contains also a term
    proportional to $(p_t + p_{\bar{t}})^{\mu}$, but this term gives
    vanishing contribution in the limit of zero electron mass.}

On the other hand, we assume the standard form for the
electron-positron couplings:
\begin{eqnarray}
&&{\mit\Gamma}_{\gamma e\bar{e}}^{\mu}
=-e\,\bar{v}(p_{\bar{e}})\gamma^{\mu}u(p_e),   \label{eq:eeg} \\
&&{\mit\Gamma}_{Ze\bar{e}}^{\mu}=\frac{g}{4 \cos\theta_W}
\bar{v}(p_{\bar{e}})
\gamma^{\mu}(v_e + \gamma _5 )u(p_e),   \label{eq:eez}
\end{eqnarray}
where $v_e \equiv -1 + 4 \sin^2\theta_W$.

Now, using the above vertices and propagators $d_{\gamma}=1/s,\ 
d_Z=1/(s-M_Z^2)$, we can represent the invariant amplitude for $e^+
e^-\to t\bar{t}$ as follows:
\begin{eqnarray}
&&M(e\bar{e} \rightarrow t\bar{t})
  =g_{\mu\nu}(d_\gamma{\mit\Gamma}^\mu_{\gamma e\bar{e}}
  {\mit\Gamma}^\nu_{\gamma t\bar{t}}+d_Z{\mit\Gamma}^\mu_{Ze\bar{e}}
  {\mit\Gamma}^\nu_{Zt\bar{t}})                               \non \\
&&\phantom{M(e\bar{e} \rightarrow t\bar{t})}
  =C_{VV}\:[\,\bar{v}_e \gamma_\mu u_e
       \cdot\bar{u}_t \gamma^\mu v_{\bar{t}} \,]
  +C_{V\!A}\:[\,\bar{v}_e \gamma_\mu u_e
       \cdot\bar{u}_t \gamma_5 \gamma^\mu v_t \,]             \non \\
&&\phantom{M(e\bar{e} \rightarrow t\bar{t})}
  +C_{AV}\:[\,\bar{v}_e \gamma_5 \gamma_\mu u_e
       \cdot\bar{u}_t \gamma^\mu v_t \,]
  +C_{AA}\:[\,\bar{v}_e \gamma_5 \gamma_\mu u_e
       \cdot\bar{u}_t \gamma_5 \gamma^\mu v_t \,]             \non \\
&&\phantom{M(e\bar{e} \rightarrow t\bar{t})}
  +C_{V\!S}\:[\,\bar{v}_e \sla{q} u_e
       \cdot\bar{u}_t v_t \,]
  +C_{V\!P}\:[\,\bar{v}_e \sla{q} u_e
       \cdot\bar{u}_t \gamma_5 v_t \,]                        \non \\
&&\phantom{M(e\bar{e} \rightarrow t\bar{t})}
   +C_{AS}\:[\,\bar{v}_e \gamma_5 \sla{q} u_e
       \cdot\bar{u}_t v_t \,]
   +C_{AP}\:[\,\bar{v}_e \gamma_5 \sla{q} u_e
       \cdot\bar{u}_t \gamma_5 v_t \,],
\end{eqnarray}
where
\begin{eqnarray*}
&&C_{VV}=-\frac{ge}{2s}[\:(A_\gamma+\delta\!A_\gamma)
         - v_e d'(A_Z+\delta\!A_Z)\:],                             \\
&&C_{V\!A}=-\frac{ge}{2s}[\:\delta\!B_\gamma
         - v_e d'(B_Z+\delta\!B_Z)\:],\\
&&C_{AV}= -\frac{ge}{2s}\: d'(A_Z+\delta\!A_Z),                    \\
&&C_{AA}=-\frac{ge}{2s}\:d'(B_Z+\delta\!B_Z),                      \\
&&C_{V\!S}=-\frac{ge}{4m_t s}
        [\:\delta C_\gamma-v_e d'\delta C_Z\:],                    \\
&&C_{V\!P}= \frac{ge}{4m_t s}
        [\:\delta\!D_\gamma - v_e d'\delta\!D_Z\:],                \\
&&C_{AS}=-\frac{ge}{4m_t s}\:d'\delta C_Z,                         \\
&&C_{AP}=\frac{ge}{4m_t s}\:d'\delta\!D_Z,
\end{eqnarray*}
with
$$
d'\equiv\frac{s}{4\sin\theta_W \cos\theta_W}d_Z\,.
$$
A straightforward calculation leads to the following differential
cross section in which $t$ and $\bar{t}$ have spins $s_+$ and $s_-$
respectively:
\begin{eqnarray}
&&\frac{d\sigma}{d{\mit\Omega}}(e^+e^-\to t(s_+)\bar{t}(s_-))  \non\\
&&=\frac{3\beta\alpha^2}{16 s^3}\:\Bigl[
\:\:D_V\:[\:\{ 4m_t^2s+(lq)^2 \}(1-s_{+}s_{-})+s^2(1+s_{+}s_{-})
                                                               \non\\
&&\ \ \ \lspace +2s(ls_{+}\;ls_{-}-Ps_{+}\;Ps_{-})
           +\:2\,lq(ls_{+}\;Ps_{-}-ls_{-}\;Ps_{+})\:]          \non\\
&&\ \ \ +\:D_A\:[\:(lq)^2(1+s_{+}s_{-})-(4m_t^2s-s^2)(1-s_{+}s_{-})
                                                               \non\\
&&\ \ \ \lspace -2(s-4m_t^2)(ls_{+}\;ls_{-}-Ps_{+}\;Ps_{-})
           -\:2\,lq(ls_{+}\;Ps_{-}-ls_{-}\;Ps_{+})\:]          \non\\
&&\ \ \
-4\:{\rm Re}(D_{V\!\!A})\:m_t\,[\:s(\psp-\psm)+lq(\lsp+\lsm)\:]\non\\
&&\ \ \ +2\:{\rm Im}(D_{V\!\!A})\:[\:lq\,\eps(s_+,s_-,q,l)
+\lsm\eps(s_+,P,q,l)+\lsp\eps(s_-,P,q,l)\:]                    \non\\
&&\ \ \ +4\:E_V\:\m s(\lsp+\lsm)+4\:E_A\:\m\,lq(\psp-\psm)     \non\\
&&\ \ \
+4\:{\rm Re}(E_{V\!\!A})\:[\:2\mq(\lsp\;\psm-\lsm\;\psp)-lq\:s\:]
                                                               \non\\
&&\ \ \
+4\:{\rm Im}(E_{V\!\!A})\:\m[\:\eps(s_+,P,q,l)+\eps(s_-,P,q,l)\:]
                                                               \non\\
&&\ \ \
+D_S\frac1{m_t^2}
[\:(lq)^2+4m_t^2 s-s^2\:][\:(4m_t^2 -s)(1-s_+ s_-)-2Ps_+ Ps_-] \non\\
&&\ \ \
-D_P\frac1{m_t^2}[\:(lq)^2+4m_t^2 s-s^2\:]
[\:s(1+s_+ s_-)-2Ps_+ Ps_-]                                    \non\\
&&\ \ \
+4\:{\rm Re}(D_{S\!P})
\frac1{m_t}[\:(lq)^2+4m_t^2 s-s^2\:](Ps_+ +Ps_-)               \non\\
&&\ \ \
+2\:{\rm Im}(D_{S\!P})
\frac1{m_t^2}[\:(lq)^2+4m_t^2 s-s^2\:]\eps(s_+,s_-,P,q)        \non\\
&&\ \ \ -\:{\rm Re}(F_1)\:\mr[\:lq\;s(\lsp-\lsm)
-\{(lq)^2+4\mq s\}(\psp+\psm)\:]                               \non\\
&&\ \ \ +2\:{\rm Im}(F_1)\:[\:s\,\eps(s_+,s_-,P,q)
+lq\,\eps(s_+,s_-,P,l)\:]                                      \non\\
&&\ \ \ +2\:{\rm Re}(F_2)\:s(\psp\;\lsm+\psm\;\lsp)            \non\\
&&\ \ \ -\:{\rm Im}(F_2)\:\frac{s}{m_t}
[\:\eps(s_+,P,q,l)-\eps(s_-,P,q,l)\:]                          \non\\
&&\ \ \ -2\:{\rm Re}(F_3)\:lq(\psp\;\lsm+\psm\;\lsp)           \non\\
&&\ \ \ +\:{\rm Im}(F_3)\:\frac{lq}{m_t}
[\:\eps(s_+,P,q,l)-\eps(s_-,P,q,l)\:]                          \non\\
&&\ \ \ -\:{\rm Re}(F_4)\:\frac{s}{m_t}
[\:lq\;(\psp+\psm)-(s-4\mq)(\lsp-\lsm)\:]                      \non\\
&&\ \ \ -2\:{\rm Im}(F_4)\:
[\:\psp\eps(s_-,P,q,l)+\psm\eps(s_+,P,q,l)\:]                  \non\\
&&\ \ \ +2\:{\rm Re}(G_1)\:
[\:\{ 4\mq s+(lq)^2-s^2 \}(1-\sss)-2s\,\psp\psm                \non\\
&&\ \ \ \lspace +lq(\lsp\;\psm-\lsm\;\psp)\:]                  \non\\
&&\ \ \ -\:{\rm Im}(G_1)\:\frac{lq}{m_t}
[\:\eps(s_+,P,q,l)+\eps(s_-,P,q,l)\:]                          \non\\
&&\ \ \ -\:{\rm Re}(G_2)\:\frac{s}{m_t}
[\:(s-4\mq)(\lsp+\lsm)-lq\;(\psp-\psm)\:]                      \non\\
&&\ \ \ -2\:{\rm Im}(G_2)\:
[\:\psp\eps(s_-,P,q,l)-\psm\eps(s_+,P,q,l)\:]                  \non\\
&&\ \ \ -\:{\rm Re}(G_3)\:\frac{lq}{m_t}
[\:lq\;(\psp-\psm)-(s-4\mq)(\lsp+\lsm)\:]                      \non\\
&&\ \ \ -2\:{\rm Im}(G_3)\:lq\,\eps(s_+,s_-,q,l)               \non\\
&&\ \ \ +2\:{\rm Re}(G_4)\:
[\:(s-4\mq)(\psp\;\lsm-\psm\;\lsp)+2\,lq\;\psp\psm\:]          \non\\
&&\ \ \ +\:{\rm Im}(G_4)\:\mr (s-4\mq)[\:\eps(s_+,P,q,l)
+\eps(s_-,P,q,l)\:]\:\:\Bigr],
\end{eqnarray}
where $\beta\equiv\sqrt{1-4m_t^2/s}$, $P$, $q$ and $l$ are defined as
$P\equiv p_e+p_{\bar{e}}\:(=p_t+p_{\bar{t}})$, $l\equiv
p_e -p_{\bar{e}}$, $q\equiv p_t-p_{\bar{t}}$, the symbol
$\epsilon(a,b,c,d)$ means $\epsilon_{\mu\nu\rho\sigma}a^\mu b^\nu
c^\rho d^\sigma$ for $\epsilon_{0123}=+1$, and
\begin{eqnarray*}
&&D_V=(s^2/e^4)(|C_{VV}|^2+|C_{AV}|^2),\ \
  D_A=(s^2/e^4)(|C_{V\!A}|^2+|C_{AA}|^2),                          \\
&&D_{V\!A}=(s^2/e^4)(C_{VV}^* C_{V\!A}+C_{AV}^* C_{AA}),           \\
&&\\
&&E_V=2(s^2/e^4){\rm Re}(C_{AV}^* C_{VV}),\ \
  E_A=2(s^2/e^4){\rm Re}(C_{AA}^* C_{V\!A}),                       \\
&&E_{V\!A}= (s^2/e^4)(C_{VV}^* C_{AA}+C_{AV}^* C_{V\!A}),          \\
&&\\
&&D_S=m_t^2(s^2/e^4)(|C_{V\!S}|^2+|C_{AS}|^2),\ \
  D_P=m_t^2(s^2/e^4)(|C_{V\!P}|^2+|C_{AP}|^2),                     \\
&&D_{S\!P}=m_t^2(s^2/e^4)(C_{V\!S}^* C_{V\!P}+C_{AS}^* C_{AP}),    \\
&& \\
&&F_1= 2m_t (s^2/e^4)(C_{VV}^* C_{V\!P}+C_{AV}^* C_{AP}),          \\
&&F_2= 2m_t (s^2/e^4)(C_{VV}^* C_{AP}+C_{AV}^* C_{V\!P}),          \\
&&F_3= 2m_t (s^2/e^4)(C_{V\!A}^* C_{V\!P}+C_{AA}^* C_{AP}),        \\
&&F_4= 2m_t (s^2/e^4)(C_{V\!A}^* C_{AP}+C_{AA}^* C_{V\!P}),        \\
&& \\
&&G_1= 2m_t (s^2/e^4)(C_{VV}^* C_{V\!S}+C_{AV}^* C_{AS}),          \\
&&G_2= 2m_t (s^2/e^4)(C_{VV}^* C_{AS}+C_{AV}^* C_{V\!S}),          \\
&&G_3= 2m_t (s^2/e^4)(C_{V\!A}^* C_{V\!S}+C_{AA}^* C_{AS}),        \\
&&G_4= 2m_t (s^2/e^4)(C_{V\!A}^* C_{AS}+C_{AA}^* C_{V\!S}).
\end{eqnarray*}
$D_{S,P,S\!P}$ are coefficients which are usually neglected and the
other coefficients are defined the same way as in ref.\cite{GH}.

The asymmetry $\delta$ can be calculated by using the above
differential cross section:
\begin{equation}
\delta=\frac{-2 \beta\, {\rm{Re}} [F_1 - \beta ^2 (s/m_t^2)D_{S\!P}]}
 {(3-\beta ^2)D_V+ 2\beta ^2 D_A - 2\beta^2 {\rm{Re}}(G_1)
 +\beta^2(s/m_t^2)(\beta^2 D_S+D_P)}.  \label{eq:cpasy}
\end{equation}
If we keep only $D_V$, $D_A$ and $F_1$ terms, this $\delta$ agrees
with the one calculated in \cite{CKP}--\cite{AS}. We find that the
term including $D_{S\!P}$ in the numerator, which consists of $\delta
C_v$ and $\delta\!D_v$, is proportional to $s$. This means there is a
possibility that $D_{S\!P}$ contributes greatly to $C\!P$ violation
at very high energy, depending on the size of $\delta C_v$ and
$\delta\!D_v$.

\begin{figure}[t]
\parbox[b]{0.46\textwidth}
{
  \begin{center}
    \epsfxsize=7cm
    \epsfbox{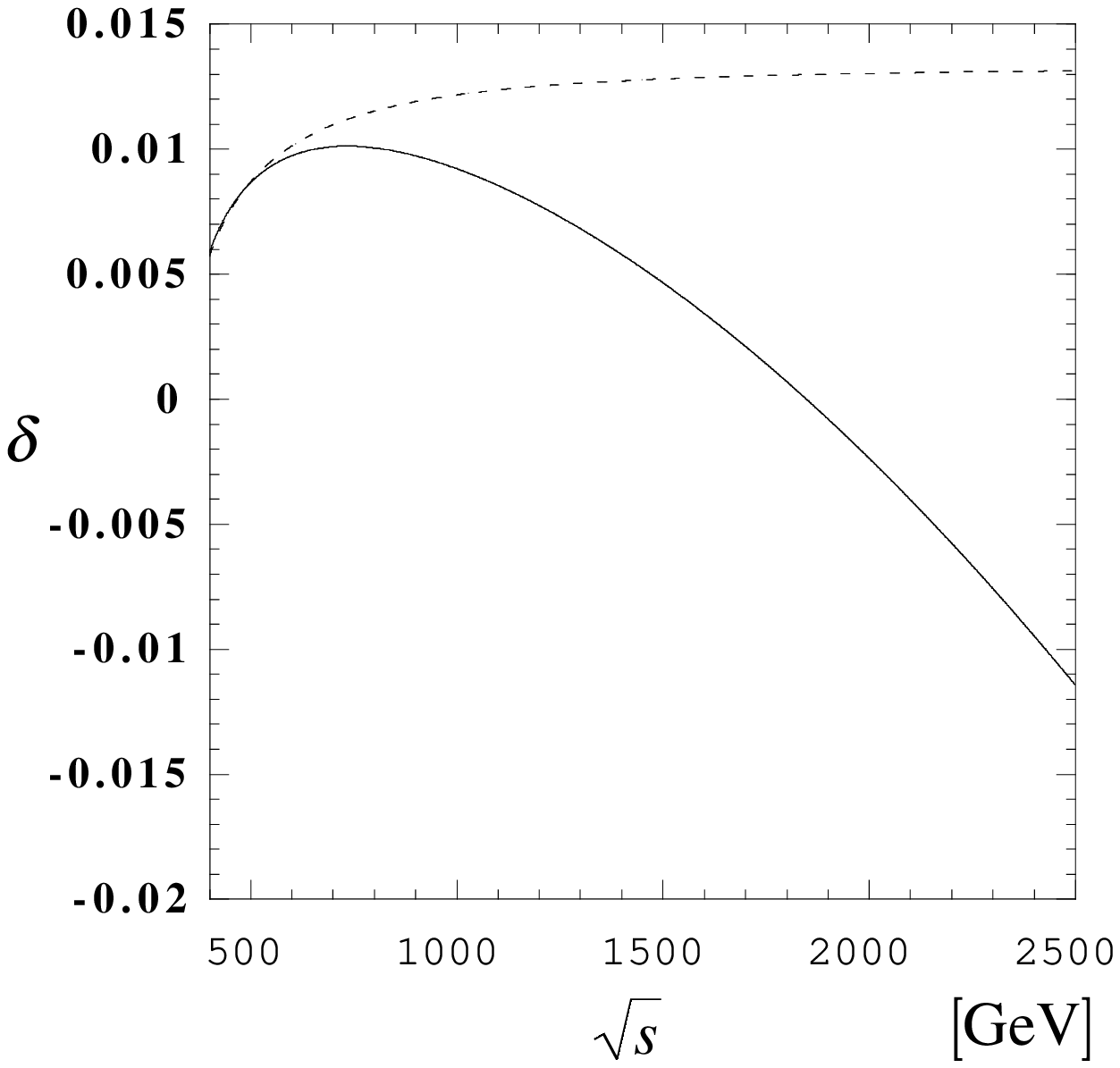}
  \end{center}
  \vspace*{-1cm}
  \caption{$C\!P$-violation asymmetry $\delta$ via our calculations
         (solid line) and the usual calculations (dotted line)
         assuming all the non-SM parameters to be 0.01.
} \label{fig:0.01}
} \hfill
\parbox[b]{0.46\textwidth}
{
  \begin{center}
    \epsfxsize=7cm
    \epsfbox{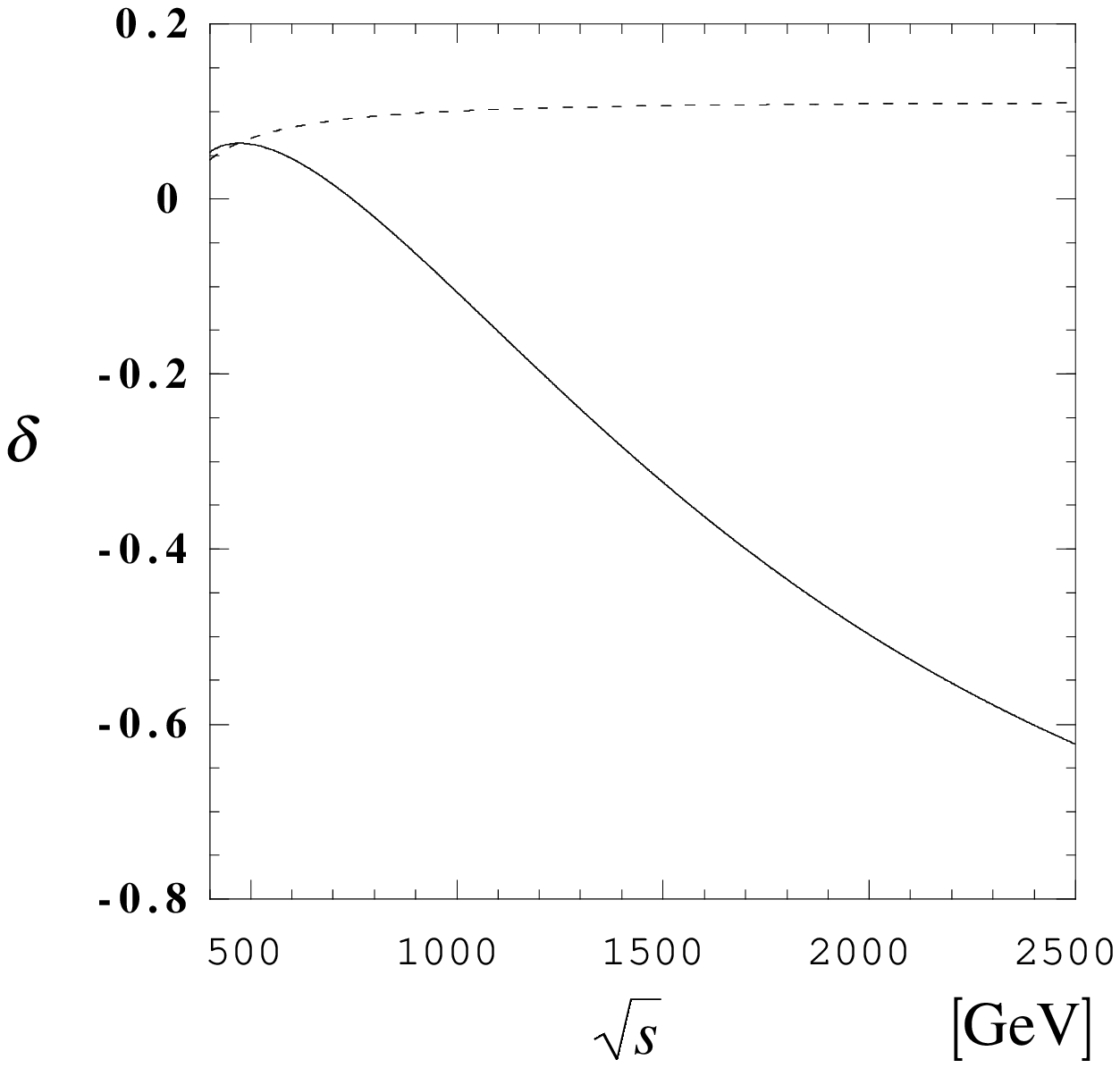}
  \end{center}
  \vspace*{-1cm}
  \caption{$C\!P$-violation asymmetry $\delta$ via our calculations
         (solid line) and the usual calculations (dotted line)
         assuming all the non-SM parameters to be 0.1.
}  \label{fig:0.1}
}
\end{figure}

Let us show the difference between our calculations and usual
calculations visually in Fig.\ref{fig:0.01} and Fig.\ref{fig:0.1},
where we use as input data $m_t=173.9$ GeV \cite{mt}, $M_Z=91.1867$
GeV \cite{data97} and $\sin^2\theta_W=0.2315$ \cite{data97}, and
assume as examples all the real and imaginary parts of $\delta\!A_v,
\cdots,\delta\!D_v$ in eq.(\ref{ff}) to be 0.01 in Fig.\ref{fig:0.01}
and 0.1 in Fig.\ref{fig:0.1}. These figures show there is no
difference around $\sqrt{s}=500$ GeV, but our $\delta$ decreases
rapidly for higher $\sqrt{s}$.   

Our calculations here are exact as long as we can treat the anomalous
couplings as constant parameters. Some comments are necessary,
however, on this point before going to summary. If these couplings
come from some new physics at a higher-energy scale ${\mit\Lambda}$,
then our results are applicable only for $\sqrt{s}<{\mit\Lambda}$.
One plausible way to estimate this ${\mit\Lambda}$ will be given by
the effective-lagrangian approach \cite{bw}. According to it, $\delta
C_v$ and $\delta\!D_v$ are both $O(m_t^2/g{\mit\Lambda}^2)$
\cite{GH2}, which leads to ${\mit\Lambda}\sim m_t/\sqrt{g\,\delta_v}$
($\delta_v=\delta C_v$ or $\delta\!D_v$) and roughly 2.5 TeV (0.8
TeV) for $\delta_v=$0.01 (0.1). Therefore our results, especially the
one in Fig.\ref{fig:0.1} must be considerably affected if this
approach describes the nature correctly, but in that case all the
other usual calculations also lose their validity anyway.

To summary, we studied contribution of the products of non-standard
parameters, which are usually neglected, to $C\!P$ violation in $e^+
e^-\to t\bar{t}$ assuming the most general $t\bar{t}\gamma/Z$
couplings. We showed there is a possibility that usual approximate
calculations may fail to give accurate results at very high
$\sqrt{s}$. We considered top productions at NLC in this paper, but
the same discussion holds also for those at hadron colliders
\cite{Hadron} if we replace $e^+$ and $e^-$ with the light quarks
(and add the gluon-fusion diagram). Finally, we cannot detect $t
\bar{t}$ directly in actual experiments. If there are no or only tiny
anomalous terms in $tbW$ couplings, it is easy to derive, e.g., the
final-lepton-energy distributions \cite{CKP,AS,GH}. In order to study
new-physics effects consistently, however, we should study the decay
process the same way as we did here for the production process. We
would like to do it elsewhere.

\vskip 0.5cm

\end{document}